\documentclass[prb,showpacs,showkeys]{revtex4}

\usepackage{graphicx}
\usepackage{dcolumn}
\usepackage{bm}

\begin{document}

\title{Topological Gravity, the Hierarchy Problem and Axion Physics}%
\author{Carlos Alberto Santos Almeida}
\email{carlos@fisica.ufc.br}
\author{Makarius Oliveira Tahim}
\email{mktahim@fisica.ufc.br} \affiliation{Departamento de
F\'{\i}sica, Universidade Federal do
Cear\'a, Caixa Postal 6030, 60455-760, Fortaleza, Cear\'a, Brazil}%


\begin{abstract}
In the last years higher dimensional physics has won importance.
Despite the Superstrings, higher dimensional effects, in measurable
scales of energy (some TeV), became only possible with
Randall-Sundrum's models (RS). In particular, recent studies in
neutrino and axion physics have proposed new and interesting
questions about neutrino mixings and new scales intermediating the
Weak and Planck scales. In this work we discuss field theoretic
models that in some aspects are similar to the RS models. Indeed,
our models contain domain walls, solitonic-like objects that mimics
the branes of the RS models. Applications are discussed ranging from
topological field theories in higher dimensions until models
containing  $D=5$ space-time torsion in the RS scenario. In
particular, we talk about subjects related to topological gravity,
the hierarchy problem and axion physics. The topological terms
studied are generalizations for $D>4$ of the axion-foton coupling in
$D=4$. Such procedure involves naturally the Kalb-Ramond field. By
dimensional reductions we obtain topological terms of the $B\wedge
F$ type in $D=4$ Chern-Simons-like and $B\wedge\partial\varphi$ type
both in $D=3$.
\end{abstract}
\pacs{11.10.Kk, 04.50.+h, 12.60.-i, 04.20.Gz}

\keywords{Randall-Sundrum scenario; Field localization; Topological
gravity; Equivalent models.}

\maketitle

\section{INTRODUCTION}
\subsection{The hierarchy problem}

May the Standard Model be placed in form of the recent insights
coming from String Theories, where several dimensions appear so
naturally? The standard model for strong, weak and electromagnetic
interactions, described by the gauge group $SU(3)\times SU(2)\times
U(1)$, has its success strongly based on experimental evidences.
However, it has several serious theoretical drawbacks suggesting the
existence of new and unexpected physical facts beyond those
discussed in the last years. One of these problems is the so called
\textit{gauge hierarchy problem} which is related to the weak
($M_{ew}$) and Planck ($M_{pl}$) scales, the fundamental scales of
the model. The central idea of this problem is to explain the
smallness of the hierarchy $M_{ew}/ M_{pl}\sim 10^{-17}$. In the
context of the minimal standard model,  this hierarchy of scales is
unnatural since it requires a fine-tuning order by order in the
perturbation theory. The first attempts to solve this problem were
the technicolor scenario \cite{A} and the low energy supersymmetry
\cite{B}. We mention that electroweak interactions have been proved
at distances $~M_{ew}^{-1}$, but gravity has only accurately
measured in the $~1cm$ range. Note that the Planck length is
$~10^{-33}cm$.

\subsection{Extra dimensions - Randall-Sundrum scenario}

With the string theories, the search of many-dimensional theories
became important. The basic idea is that extra dimensions can be
used to solve the hierarchy problem: the fields of the standard
model must be confined to a $(3+1)$-dimensional subspace, embedded
in a $n$-dimensional manifold. In the seminal works of Arkani-Hamed,
Dimopoulos, Dvali and Antoniadis \cite{C}, the $4$-dimensional
Planck mass is related to $M$, the fundamental scale of the theory,
by the extra-dimensions geometry. Through the Gauss law, they have
found $M^{2}_{pl}=M^{n+2} V_{n}$, where $V_{n}$ is the extra
dimensions volume. If $V_{n}$ is large enough, $M$ can be of the
order of the weak scale. However, unless there are many extra
dimensions, a new hierarchy is introduced between the
compactification scale, $\mu_{c}= V^{-\frac{1}{n}}$, and $M$. An
important feature of this model is that the space-time metric is
factorizable, i.e., the $n$-dimensional space-time manifold is
approximately a product of a $3$-dimensional space by a compact
$(n-3)$-dimensional manifold.

Because of this new hierarchy, Randall and Sundrum  \cite{D} have
proposed a higher dimensional scenario that does not require large
extra dimensions, neither the supposition of a metric factorizable
manifold. Working with a single $S^{1}/Z_{2}$ orbifold extra
dimension, with three-branes of opposite tensions localized on the
fixed points of the orbifold and with adequate cosmological
constants as $5$-dimensional sources of gravity, they have shown
that the space-time metric of this model contains a redshift factor
which depends exponentially on the radius $r_{c}$ of the
compactified dimension:
\begin{equation}\label{eq1}
d s^{2}= e^{-2k r_{c}|\phi|}\eta_{\mu\nu} d x^{\mu}d x^{\nu}-r_{c}
d\phi^{2},
\end{equation}
where $k$ is a parameter of the order of $M$, $x^{\mu}$ are Lorentz
coordinates on the surfaces of constant $\phi$, and
$-\pi\leq\phi\leq\pi$ with $(x,\phi)$ and $(x,-\phi)$ identified.
The two $3$-branes are localized on $\phi=\pi$ and $\phi=0$. In
fact, this scenario is well known in the context of string theory.
The non-factorizable geometry showed in Eq.(\ref{eq1}) has important
consequences. In particular, the $4$-dimensional Planck mass is
given in terms of the fundamental scale $M$ by
\begin{equation}\label{eq2}
M_{pl}^{2}=\frac{M^{3}}{k}[1-e^{-2k r_{c}\pi}],
\end{equation}
in such a way that, even for large $k r_{c}$, $M_{pl}$ is of the
order of $M$. The other success of this scenario is that the
standard model particle masses are scaled by the warp exponential
factor.

\subsection{Topological gravity: some motivations}

Despite these important developments in classical Einstein gravity,
background independent theories are welcome. As an example it is
worth mentioning the Quantum Loop Gravity, developed mainly by
Asthekar et al \cite{E}. Also the problem of background dependence
of string field theory has not been successfully addressed. The
string field theory has a theoretical problem: it is only
consistently quantized in particular backgrounds, which means that
we have to specify a metric background in order to write down the
field equations of the theory. This problem is fundamental because a
unified description of all string backgrounds would make possible to
answer questions about the selection of particular string vacua and
in general to give us a more complete understanding of geometrical
aspects of string theory \cite{F}.

Due to these developments, we regard as an important subject look
for topological theories in the brane context, with the major
objective to add quantum information to the Randall-Sundrum
scenario. In this sense, as the first step, we study topological
theories on brane-worlds in several dimensions. In this part of the
work we construct topological theories in brane-worlds. The
brane-world is regarded as a kink-like soliton and it appears due to
a spontaneous symmetry breaking of a Peccei-Quinn-like symmetry.
Topological terms are obtained by generalizing to many dimensions
the axion-foton anomalous interaction.

\section{TOPOLOGICAL TERMS IN BRANE-WORLDS}

We implement the theory through the following action in $D=5+1$:
\begin{equation}
S=\int d^{6}x\left( -\frac{1}{2\left( 3!\right) }H_{MNP
}H^{MNP}+g\varepsilon ^{MNPQRS}\phi
\left( z\right) H_{MNP}H_{QRS}+\frac{1}{2}%
\partial _{M }\phi \partial ^{M}\phi +V\left( \phi \right)
\right)   \label{eq1-1}
\end{equation}
In this action, $H_{MNP}=\partial _{M}B_{NP }+\partial _{N }B_{PM
}+\partial _{P}B_{MN }\left(M,N ,...=0,...,5\right) $ is the field
strength of an antisymmetric tensor gauge field $B_{MN}$. The field
$B_{\mu \nu }$ has an important function in string theory: it
couples correctly with the string world-sheet in a very similar way
to the coupling of a gauge vector field $A_{M }$ to the universe
line of a point particle. The field $\phi $ is a real scalar field,
and $V\left( \phi\right) =\lambda \left( 1-\cos \phi\right)$, is a
potential that provides a phase transition.

The second term in the action (\ref{eq1-1}) is a term that
generalizes the coupling  that appears from the anomaly of the
Peccei-Quinn quasisymmetry in $D=3+1$, namely,
$\phi\rightarrow\phi+a$. For such, the space-time dimension is
$D=5+1$ and the hypersurface is a $D=4+1$ world. Now we can work
with the second term of (\ref{eq1-1}) (considering that $\phi$ only
depends of the z-coordinate $z$) in order to obtain new terms by
integration by parts. Considering that the $B_{MN}$ field weakly
depends on the z-coordinate, we obtain:
\begin{equation}
S_{top.}=\int d^{5}x\left( k\varepsilon ^{MNPQ R }B_{MN}H_{PQR
}\right) \label{eq6-1}
\end{equation}
This last equation shows that over the hypersurface an effective
topological term appears with a coupling constant \textbf{k} that
have canonical dimension of mass. The theory over the hypersurface
is completely five-dimensional. This term is very similar to the
Chern-Simons term, that is written in $D=2+1$ with a gauge vector
field $A_{\mu }$. Nevertheless, the term (\ref{eq6-1}) is written
only with tensorial antisymmetric fields $B_{\mu \nu }$. Such term
have been used to explain some peculiarities of the Cosmic Microwave
Background Radiation (CMBR) within the Randall-Sundrum scenario
\cite{indianos}. It is interesting now to observe the properties of
the action (\ref{eq1-1}) in lower dimensional space-times using
dimensional reduction. Thus, supposing that the fields of the action
(\ref{eq1-1}) are independent of the coordinate $x_{M }\equiv x_{5}$
which is not the argument of the field $\phi\left( z\right)$, we
find a new action in $D=4+1$. This action has now a vectorial gauge
field reminiscent of the reduction, and contains yet the real scalar
field $\phi$, that again may give rise to the formation of a lower
dimensional domain wall-brane. In this case, the space-time
dimension is $D=4+1$ and the hypersurface is a $D=3+1$ universe. If
we observe the theory over the solitonic hypersurface we will obtain
that,
\begin{equation}
S_{top.}=k \int d^{4}x\left( \varepsilon ^{4\nu \alpha \rho \sigma
}V_{\nu \alpha }B_{\rho \sigma }\right)   \label{eq10-1}
\end{equation}
If the field $V_{\mu }$ is identified with the potential four-vector
$A_{\mu }$ then we obtain the action for the $B\wedge F$ model on
the domain wall-brane. This action, under certain conditions, can
give rise to a mechanism of topological mass generation for the
field $A_{\mu }$ or for the field $B_{\mu \nu }$.

The discussion for lower dimensions ($D=3+1$) and ($D=2+1$), using
the same methods, will lead us to the following topological action:
\begin{equation}
S_{top.}=k \int d^{4}x\left[ \varepsilon ^{\mu \nu \alpha \rho }\phi
\left( z\right) \partial _{\mu }\phi \partial _{\nu }B_{\alpha \rho
}+\varepsilon ^{\mu \nu \alpha \rho }\phi \left( z\right) F_{\mu \nu
}W_{\alpha \rho }\right]   \label{eq12-1}
\end{equation}
The fields $\phi$ and $W_{\alpha \rho }=\partial _{\alpha }W_{\rho
}-\partial _{\rho }W_{\alpha }$ emerge as degrees of freedom
reminiscent of the reduction. If we work with the first term of
(\ref{eq12-1}) on the domain wall, we will find a different
topological theory:
\begin{equation}
S=\int d^{3}x\left( g\varepsilon ^{abc}\partial _{a }\phi
B_{bc}\right)   \label{eq13-1}
\end{equation}
Identifying again in (\ref{eq12-1}), in the second term, the vector
field $W_{\mu }$ as the gauge field $A_{\mu }$, then we will obtain
the anomalous interaction term between the real scalar field $\phi$
and the field $A_{\mu }$ . This term, rearranged on the domain wall,
reduces to the Chern-Simons term.

\section{TOPOLOGICAL APPROACH TO THE HIERARCHY PROBLEM}

In this section we review an alternative to the central point of the
Randall-Sundrum model \cite{H}, namely, the particular
nonfactorizable metric. Using a topological theory, we show that the
exponential factor, crucial in the Randall-Sundrum model, appears in
our approach, only due to the brane existence instead of a special
metric background. In order to study the hierarchy problem we choose
to work with topological gravity. Motivated by current searches in
the quantum gravity context, we study topological gravity of
$B\wedge F$ type. Then, we can affirm that our model is purely
topological because $1)$ the brane exists due to the topology of the
parameter space of the model and $2)$ gravity is metric independent.
We will see that these features give us interesting results when
compared to the Randall-Sundrum model.

\subsection{The Model}
The model is based on the following action:
\begin{equation}\label{eq3}
S= \int d^{5} x [\frac{1}{2}\partial_{M}\phi\partial^{M}\phi+
k\varepsilon^{MNPQR}\phi H_{MNP}^{a} F_{QR}^{a}-V(\phi)].
\end{equation}
In this action the $\phi$ field is a real scalar field that is
related to the domain wall. The fields $H_{MNP}^{a}$ and
$F_{QR}^{a}$ are non-abelian gauge fields strengths and will be
related to the gravitational degrees of freedom. Namely, in pure
gauge theory,
$H^{a}_{MNP}=\partial_{M}B^{a}_{NP}-\partial_{N}B^{a}_{PM}-\partial_{P}B^{a}_{MN}+gf^{abc}A^{b}_{M}B^{c}_{NP}$
and
$F^{a}_{MN}=\partial_{M}A^{a}_{N}-\partial_{N}A^{a}_{M}+g'f^{abc}A^{b}_{M}A^{c}_{N}$.
The second term of this action is a topological version of the terms
studied above. The action (\ref{eq3}) is invariant under a
Peccei-Quinn  symmetry transformation $\phi\rightarrow\phi+2\pi n$.
The potential is
\begin{equation}\label{eq4}
V(\phi)=\lambda(1-\cos\phi),
\end{equation}
which preserves the Peccei-Quinn symmetry. Nevertheless, it is
spontaneously broken in scales of the order of $M_{PQ}\sim
10^{10}-10^{12}$GeV. We propose the following potential
\begin{equation}\label{eq5}
V(\phi)=\frac{\lambda}{4}(\phi^{2}-v^{2})^{2},
\end{equation}
which explicitly breaks the $U_{PQ}(1)$ Peccei-Quinn symmetry, in
order to generate a brane in an energy close to the weak scale. With
this particular choice of the potential, the existence of the brane
is put on more consistent grounds. In other words, the brane appears
almost exactly in an energy scale of the universe near the symmetry
breaking scale of the electroweak theory. This feature was assumed
in previous works without a careful justification. However, this
mechanism leads to a large disparity between the Planck mass
$M_{PL}\sim 10^{18} GeV$ and the scale of explicit breaking of
$U_{PQ}(1)$ which is relatively close to the weak scale, $M_{ew}\sim
10^{3} GeV$: we assume this disparity as a new version of the
hierarchy problem. Consider now the solution
$\phi(x_{4})=v\tanh(\sqrt{\frac{\lambda}{2}}v x_{4})$.
This solution defines a $3$-brane embedded in a $(4+1)$-dimensional
space-time. The mass scale of this model is $m=\sqrt{\lambda}v$ and
the domain wall-brane thickness is $m^{-1}$. With this information
we can now discuss the effective theory on the domain wall-brane. An
integration by parts of the topological term in the action
(\ref{eq3}) will result in
\begin{equation}\label{eq9}
S\sim \int d^{4} x \varepsilon_{\nu\alpha\rho\lambda}
B_{\nu\alpha}^{a} F_{\rho\lambda}^{a} [\lim_{r_{c}\rightarrow
+\infty} k'\int_{0}^{r_{c}} d x_{4} \partial_{4}\phi(x_{4})],
\end{equation}
where $r_{c}$ represents the extra dimension. This last conclusion
denotes the domain wall-brane contribution to the effective
four-dimensional theory. We can see that, effectively on the domain
wall-brane, the theory is purely $4$-dimensional (this is important)
and is described by a non-abelian topological $B\wedge F$ term. It
can be shown that, under parameterizations by tetrad fields, a
$B\wedge F$ type action gives us
\begin{equation}\label{eq9a}
\int d^{4} x k\varepsilon^{\nu\alpha\rho\lambda} B_{\nu\alpha}^{a}
F_{\rho\lambda}^{a}\rightarrow k\int d^{4} x \sqrt{g}R,
\end{equation}
which is the Einstein-Hilbert action for the gravitational field,
where $R$ is the scalar curvature and $g$ stands for the space-time
metric. From Eqs. (\ref{eq9}) and (\ref{eq9a}), we can see the
relation between the Planck mass $k_{4}$ in $D=4$ and the extra
dimension:
\begin{equation}\label{eq10}
k_{4}=\lim_{r_{c}\rightarrow +\infty} k'\int_{0}^{r_{c}} d x_{4}
\partial_{4}\phi(x_{4}).
\end{equation}
The limit $r_{c}\rightarrow +\infty$ ensures the topological
stability of the domain wall-brane. By the substitution of the
aforementioned solution $\phi(x_{4})$ in Eq. (\ref{eq10}),
considering a finite $r_{c}$ (which means that the domain wall-brane
is a finite object), we can show that
\begin{equation}
k_{4}=k'v(1-e^{-2y}) (1+e^{-2y})^{-1}\label{eq11},
\end{equation}
where $y=\sqrt{\frac{\lambda}{2}}v r_{c}$ is the scaled extra
dimension. This result is very interesting: as our model is a
topological one, the exponential factor must not appear from any
special metric. Here, the exponential factor appears only due to the
domain wall-brane existence. As in the Randall-Sundrum model, even
for the large limit $r_{c}\rightarrow +\infty$, the $4$-dimensional
Planck mass has a specific value. This is the reason why we believe
that our approach can be useful to treat the hierarchy problem. It
is possible to obtain scaled masses to the confined matter using
zero modes attached to the domain wall-brane. We address the fact
that we only use the domain wall-brane characteristics.

\section{ON AXION PHYSICS: some perspectives}

In this section we make a brief review of recent developments about
axion physics and propose some new perspectives in this area. The
motivations for these discussions are only mathematical: the
topological terms studied in the sections above can give us some
insights. Extra dimension physics gives new viewpoints for axion
studies. In particular, an interesting problem is how to generate
the axion scale ($f_{a}=10^{10}\sim 10^{12}$ Gev) that is
intermediate to the Planck scale ($M_{p}\approx 10^{18}$ Gev) and to
the electroweak symmetry breaking scale ($M_{ew}\approx 1$ Gev).
Such problem may be solved naturally by a $5D$ orbifold field
theories \cite{I} if the axion originates from a higher-dimensional
parity-odd field $C_{M}$:
\begin{equation}
S_{5D}=\int d^{4}x dy \sqrt{-G}
(\frac{1}{l^{2}}C_{MN}^{2}+\frac{\kappa}{\sqrt{-G}}\epsilon^{MNPQR}C_{M}
F_{NP}^{a} F_{QR}^{a}+\ldots)
\end{equation}
The action contains a Chern-Simons coupling to the $U(1)$ gauge
field $C_{M}$; $F_{NP}^{a}$ is a standard model non-abelian field
strength. The axion appears as the $C_{5}$ field, a component of
$C_{M}$, together with the correct scale $f_{a}$.

On the other hand, interesting studies have been made in gauge
fields localization procedures. For the brane of the
$\delta$-function type, Dvali et al. \cite{J} have shown that, for
gauge fields, localization holds for specific distances but there is
an effect of dissipation of cosmic radiation to extra dimensions for
large distances. The lagrangian is the following:
\begin{equation}
L=-\frac{1}{4g^{2}}F_{AB}^{2}-\frac{1}{4e^{2}}F_{\mu\nu}^{2}\delta
(y)+\ldots
\end{equation}
The first term refers to the bulk physics while the second one, to
the brane physics ($A,B=1,\ldots ,5$ and $\mu,\nu =1,\ldots ,4$). On
the brane, the $F^{2}$ term can be generated by radiative
corrections due to localized matter fields. An interesting problem
is to analyze axion physics in this context in order to see if the
same phenomenon happens for axions. In $D=4$, the axion field may be
described by an antisymmetric field $B_{\mu\nu}$. A nice way to
study this question would be to start with the following model in
$D=5$:
\begin{equation}
L=H_{ABC}^{2}+H_{\mu\nu\alpha}^{2}\delta (y)+\ldots
\end{equation}

In another way (equivalent theories on branes) we can try in $D=5$
the following model:
\begin{equation}
L=H_{ABC}^{2}-F_{AB}^{2}+\delta
(y)[\varepsilon^{\mu\nu\alpha\beta}B_{\mu\nu}F_{\alpha\beta}+A_{\mu}A^{\mu}]
\end{equation}

On the brane we have a partially topological theory that is
equivalent to a $H^{2}$ theory. In this case we have to consider
that we can get the equations of motion of the brane physics
independently of the bulk physics. We can do this if we consider low
energies in the $D=4$ world. In this case we can construct a
mechanism of axion localization on branes in a different fashion.

\section{CONCLUSIONS}

By a procedure of dimensional reduction, we have constructed several
Chern-Simons-like topological terms, in abelian and in non-abelian
theories. The domain wall-brane has been simulated by a kink-like
soliton embedded in a higher dimensional space-time and it has
emerged  due to a spontaneous symmetry breaking of a specific
discrete symmetry, namely, a Peccei-Quinn-like symmetry. We have
shown that a simple topological model in field theory has the
necessary features to solve the Gauge Hierarchy problem in a very
similar way to the one found by Randall and Sundrum. With this model
we have built a stable $3$-brane (a domain wall-brane) that
simulates our four-dimensional Universe and we have argued the
possibility of topological gravity localization. Because of these
facts, the exponential factor appears only due to the existence of
the domain wall-brane and not from a special metric. We have
discussed some of our perspectives on axion physics. We have noted
that we can construct a mechanism of axion localization using a
partially topological field theory. Studies related to generation of
axion scales will follow in a forthcoming paper.

\section*{Acknowledgments}
This work was supported in part by Conselho Nacional de
Desenvolvimento Cient\'{\i }fico e Tecnol\'{o}gico-CNPq and
Funda\c{c}\~{a}o Cearense de Amparo \`{a} Pesquisa-FUNCAP.

\end{document}